# Landau Raman Regulation Observed from Geometric Graphene Structures Including Single-Wall Carbon Nanotubes

K. P. S. S. Hembram,[1] Jeongwon Park[2] and Jae-Kap Lee[1]*

[1]Center for Opto-Electronic Materials and Devices, Korea Institute of Science and Technology (KIST), Seoul 02792, Republic of Korea

[2]Department of Electrical and Biomedical Engineering, University of Nevada, Reno, NV 89557, USA

*Corresponding author: Email: jklee@kist.re.kr (Jae-Kap Lee)*

**ABSTRACT**

Here, we report Landau regulation in Raman modes for geometric graphene structures including single-wall carbon nanotubes (SWNTs). First-principles calculations show that radial-tangential Raman modes (RTMs) (comprising radial and tangential Eigenvectors) appear in order, following radial breading mode (RBM) (~150 cm$^{-1}$). RTMs reveal two degenerated or non-degenerated frequencies per node and extend upto $G$ (~1590 cm$^{-1}$). In Raman measurements for SWNTs, we demonstrate the existence of RTMs at ~180-1440 cm$^{-1}$, which have been partially interpreted as RBM band for ~180-300 cm$^{-1}$ and intermediate frequency mode for 600-1300 cm$^{-1}$. We reinterpret the RBM band as a coalescent signature of RBM (~150-170 cm$^{-1}$) and 1$^{st}$ node RTM (~170-190 cm$^{-1}$) for geometric graphene structures of ~2.0-2.2 nm in diameter. The sequential appearance of RBM and RTM and the revelation of the double degeneracy are due to Landau regulation working on spatial variation of the radial and tangential Eigenvectors. Our findings provide a further way of understanding the nature of low dimensional materials utilizing Raman spectroscopy.

# Introduction

Low energy Raman signals below ~300 cm$^{-1}$ for single-wall carbon nanotubes (SWNTs) are additional vibrational signatures of the atoms and becomes a fingerprint to evaluate physical properties of the structure. For instance, the radial breathing mode (RBM) for single-wall carbon nanotube (SWNT), which appears in the range of ~100-300 cm$^{-1}$ (shifted by ~10 cm$^{-1}$ with laser power and frequencies),[1] has been used as a criterion to determine not only its diameter but also electronic properties.[1-5] These are due to the general understanding that the RBM band at the low energy scale is originated by synchronous radial vibration of a tubular graphene which may related with a diameter of SWNTs, affecting in turn their chirality. The low energy Raman signals appear diversely in frequency and shape (peak or band), and this diversity has been attributed to the presence of plural (bundled) SWNTs with different diameters.[1-5]

With the recent progress in synthesis of graphene (or thin graphite), on the other hand, the low energy RBM signals have been observed from graphene or nanographite structures.[6-12] This is unexpected because RBM signals have been reported to be unique to SWNTs. Lee *et al.* have shown that low energy Raman signals of graphene structures (~100-500 cm$^{-1}$) are due to the radial mode (RM) formed by end curvatures of mono- or bilayer graphene.[13] Some researchers name the low energy signals as RBM-like without further explanation and others claim them to be layer breathing modes for planar graphene structures.[6,7] Indeed, the upper limit of the low energy signals for SWNTs has been inconsistently reported to be ~200, ~250, ~300, ~350, and ~400 cm$^{-1}$, while many Raman spectra for well-defined SWNTs reveal signals distributed from ~100 cm$^{-1}$ to ~1300 cm$^{-1}$.[1,14-17] Some researchers name the signals between ~600 and ~1300 cm$^{-1}$ as intermediate-frequency Raman modes (IFM),[5,14,15] which are reported to be independent from the excitation wavelength unlike the RBM signals.[1] These indicate that the nature of the vibrational modes at a low energy Raman spectrum still puzzles in spite of numerous investigations for the different graphene structures.

Recently, there has been an issue on the structure of SWNTs, *i.e*., non-tubular (helical) structure[18,19] which is contrary to the conventional regular tube model (Fig. S1). The helix model is supported by high-resolution transmission microscope (HRTEM)[18,20-23] and scanning tunneling microscope (STM)[24,25] images where traces of graphene helix are evident (Fig. S2) as well as energy calculation.[18] Park *et. al*. also reinterpreted SWNTs to be a graphene helix by Raman analysis of hydrogenated and dehydrogenated SWNTs.[26] They showed that the typical $D$ and $G^-$ for SWNTs are of the signature of their helical structure.

Landau theory provides information for the stability of physical structures of materials in terms of free energy, and is useful to understand their phase transition. It was extended to Landau-Ginzburg (L-G) theory which can explain many physical phenomena at certain order. L-G theory is a mathematical formalism based on series expansion concept. It was initially suggested to explain superconductivity,[27,28] and extended to understand the vertex dynamics with defects[29] and topological excitations in Josephson junction.[30] Due to its easy solvability, it was further extended to various area in quantum field theory describing certain limitation on quantum criticality,[31] optical conductivity of topological surface states with supersymmetry.[32] In particle physics, it explains the unique vacuum state and a potential energy with degenerate critical point. Looking at the similarity in phase transition in supersymmetric gauge theory, the renormalization group flow at sigma models on Calabi-Yau manifolds mimic the L-G theory.[33,34] Similar formalism is used to describe low energy dynamics of 4-dimension gauge theory with monopole and to construct brane.[35] In this paper, we explore vibrational features for regular and opened (helical) tube structures as well as curved graphene with first-principles calculation and measurement, and reveal that RBM, RTM and *G* are governed by Landau regulation.

**Results and discussion**

Raman spectra obtained from two kinds of commercial SWNTs are shown in Fig. 1. Each spectrum reveals a clear low energy band of ~100-200 cm$^{-1}$ (inset) (generally interpreted as RBM band), a *D* band at ~1350 cm$^{-1}$ and a strong *G* band at ~1590 cm$^{-1}$, which are typical to SWNTs.[1-5] Weak but clear peaks or band signals at ~200-1500 cm$^{-1}$ are evident in the zoom-in spectra (Fig. 1(b)). Fig. 2 reveals a series of active-Raman modes simulated for different geometric graphene structures, demonstrating the existence of RTMs following the RBM or the localized RBM (*l*-RBM) where the latter is for opened SWNTs or curved graphene. RTMs are mixture of radial and tangential Eigenvectors while RBM is composed of pure radial Eigenvectors. RTMs are featured by *l*-RMs (a set of RMs with a decentered focus) and appear in between RBM and *G* bands (Fig. S3). The simulation also shows that RBM and *G* bands are both extreme conditions of RTMs comprising pure RM and pure tangential mode (TM) Eigenvectors respectively. Frequencies of RBMs (*l*-RBM) and RTMs for SWNTs depend on their diameters (Fig. 2(b,c)) as well as local curvatures (Fig. 2(d,e)). With increase of node, the frequencies of RTMs monotonically increase (Fig. S4) and the *l*-RMs become weaker.

Total number of RTMs ($N_{RTM}$) for the regular tube structures is $2n$ where $n$ is a number of node and '2' represents two modes per node (Fig. 2(b,c) series). Two modes in each node for the regular tube structure are degenerate (same frequency) while those for opened tube structure are non-degenerate (different frequency). We attribute the non-degeneracy as well as the appearance of the *l*-RBM (*l*-RM) and the lack of one 1$^{st}$ node RTM for the opened tube structures to their structural asymmetry (Fig. 2(c)) where one mode becomes more energetic than the other. The locally distorted opened SWNT structure reveals strong *l*-RMs on the steep edge curvature at higher nodes up to the 4$^{th}$ (Fig. S3).

Regardless of structure of SWNTs, two types of Raman mode (*i.e.*, RBM and RTM) appear in the low energy range (below *G*) and each RTM has two degenerate or non-degenerate frequencies. The condition for RBM (or *l*-RBM) and RTM can be depicted by Landau free energy formula, $f(\delta) = f_0(\delta) + \alpha (\delta - \delta_c)m^2 + 1/2\, \beta m^4$, $\alpha > 0$, $\beta > 0$, where $f(\delta)$ is the free energy functional, $\delta$ is radial displacement of atoms, $m$ is the order parameter, $\alpha$ and $\beta$ are parameters. When $\delta > \delta_c$, the phase has one energy state revealing one minimum (RBM or *l*-RBM), while the phase has two energy states revealing two minima (*i.e.*, RTM) when $\delta < \delta_c$. RBM occurs only when all atoms in the curved graphene structures have a synchronous radial vibration, *i.e.*, $\delta > \delta_c$.

Figure 3 clearly elucidates that each RTM is a combination of RM and TM Eigenvectors where the former appears as the unique *l*-RM with a decentered focus (red dots in Fig. 3(a,b)). Each mode can be described by a correlated Eigen function (CEF), $C(\psi^1\, \psi^2..\psi^n)$ where each term is composed of RM and TM components of the atoms, *i.e.*, $\psi^i = C^i_{rm}\psi^i_{rm} + C^i_{tm}\psi^i_{tm}$ and $i$ is a natural number. RBM is the 1$^{st}$ term of CEF and can be expressed as $C^1_{rm}C^2_{rm}...C^n_{rm}\psi^1_{rm}\psi^2_{rm}...\psi^n_{rm}$, where all the Eigen vectors are directed towards the center. RTMs can be manifested from the higher order terms of CEF with spatial variations of RM and TM.

As each RTM is a combination of *l*-RM and TM, its Eigenvectors vary gradually from one to another. This spatial variation of Eigenvectors can be explained by the Landau-Ginzburg theory (see Supplementary Information), which is represented by a sine curve, $A_0 sin(\omega t+\varphi)$, where $A_0$ is the amplitude, $\omega=2\pi f$, the angular frequency (*f* is the frequency of oscillation), *t* is the time period and $\varphi$ is the phase (Fig. 3(a',b')). The sine curve is regulated by the coherence length ($\xi$), a cycle of TM-RM-TM-RM-TM variation (the distance moved by the wave per one oscillation), which is described as $2\pi r/n$ where *r* and *n* are a radius of the tubular structures and a number of node, respectively. With increase in node, the coherence length shortens and the amplitude (*A*) of the sine curve (corresponding to pure radial components of *l*-RMs) decreases. The coherence length, which is intrinsic characteristics of the geometry, also varies with the tubes dimension.

From the figures of low node RTMs featured by *l*-RMs (which consistently vary with node: Fig. 2,3), we regard the depth of focus ($d_f$) of the *l*-RMs of RTMs (Fig. 3(a)) as an indicator to estimate a variation of the intensity of RTMs (where RM is dominant) with a node. Indeed, it can be represented as a relative amplitude (*A*) (Fig. 3(a,b)) of a pure RM component of *l*-RM which enables us to estimate relative intensities of the RTMs ($I \propto A^2$). The depth of focus decreases with increase in the node where steep decrease is evident for 1st and 2nd node RTMs (Fig. 3(c)). The variation of the depth of focus with a node is consistent with that of the coherence length ($\xi$) with the node (see inset in fig. 3(c)). Here, we generalize the node dependence of the RTMs in terms of the coherence length with an equation, $2\pi r \left(1/n1 - 1/n2\right) = K\Delta v$, where the left term corresponds to the change in coherence length, and $\Delta v$ is the change in frequency with respect to the adjacent nodes and $K$ is the proportionality constant in the unit of cm$^{-2}$.

With the simulation data (Fig. 2) and the specific analysis on individual RTM (Fig. 3), we assign the typical strong peak and the shoulder peak at 169 and 188 cm$^{-1}$ for arc-SWNTs (Fig. 1) to their RBM (or *l*-RBM) and 1$^{st}$ node RTM, respectively. For Ocsial SWNTs, we assign 120, 137, 149 cm$^{-1}$ peaks to their RBMs (or *l*-RBMs) (*i.e.*, for SWNTs with different dimensions) and ~166 cm$^{-1}$ peak to 1$^{st}$ node RTM for a SWNT which reveals the 149 cm$^{-1}$ peak as RBM (or *l*-RBM), respectively. We expect that the signals of the 1$^{st}$ node RTMs for the SWNTs, which reveal 120 and 137 cm$^{-1}$ peaks as RBM (or *l*-RBM), may not appear due to strong 137 and 149 cm$^{-1}$ peaks. The analysis suggests that Ocsial SWNTs may be diverse in dimension relatively.

We also assign many signals in the zoom-in spectra (Fig. 1(b)) to RTMs; 299 and 415 cm$^{-1}$ peaks of Ocsial-SWNTs to the 2$^{nd}$ and the 3$^{rd}$ node RTMs respectively, 343 cm$^{-1}$ peak of Arc-SWNTs to the 2$^{nd}$ node RTM, and the inconsistent signals at 600-1500 cm$^{-1}$ of both samples to the higher node RTMs (>5$^{th}$) (Fig. S3). We expect that higher node RTMs near ~1500 cm$^{-1}$ (such as 1419 and 1440 cm$^{-1}$ peaks) are relatively strong due to the contribution of the TM, like the *l*-RM for lower RTMs near RBM, which explains the formation of slowly varying symmetric spectra (without *D* signal) as shown in Fig. 1(b).

From the variation of RBM and the 1st node RTM frequencies with the diameter of regular or opened SWNTs shown in Fig. 4, we estimate diameters of Ocsial-and Arc-SWNTs to be 2.0-2.2 and ~2.0 nm respectively regardless of whether their structure is regular or helical.[18] The estimated diameters are similar to those directly measured by HRTEM observation for SWNTs.[18-21]

It has been observed that graphene with curvature produces pseudo magnetic field.[36,37] Slight displacement of carbon atoms from equilibrium positions in our case leads to strain, thus producing pseudo magnetic field. We expect that the pseudo magnetic field is responsible for the different Landau like levels. Unlike the tubular structure, the helical structure possesses dangling carbon atoms at the edges which are asymmetric with respect to the axis and breaks the rotational symmetry. These further add uneven strain to the system, explaining the diverse and inconsistent RTMs (Fig. 1(b)) which have been reported as IFM.[5,14,15] With the reports for SWNTs to be a graphene helix,[18,26] we attribute the excitation wavelength dependence of the low energy Raman signals[1-3,5] to diverse curvatures of the helical edge (Fig. S3). Indeed, SWNTs reveal diverse edge structures including helical[18] or locally unrolled.[38] The SWNTs with the diverse edge curvature can be affected by environmental condition such as bundled or individual and freestanding or deposited on substrate. We also expect that the low energy Raman signals observed at 500-1100 cm$^{-1}$ from graphene structures[13] (Fig. S5) are of the RTMs for curved graphene (Fig. 2(d)).

Similar phenomena, showing well-resolved quantized Landau level, are observed in graphene structures.[39,40] The spatial variation of the RTM Eigenvectors resembles the variation of vector fields from the divergence (comparable to RBM) to anti-symmetric curl (comparable to *G*) (Fig. 2(e)). These suggest that our approach to Raman modes based on Landau theory can be extended to explain the transition of the vector field from divergence to anti-symmetric curl.

## Summary

In summary, Landau theory works on Raman modes. Geometric graphene structures including SWNTs and curved graphene reveal RTMs. They have been partially interpreted as RBM (150-300 cm$^{-1}$) and IFM (600-1300 cm$^{-1}$). The results indicate that the RBM model is not enough to explain the low energy Raman signals of SWNTs regardless of whether the structure is regular or helical. The diversity of the low energy peaks in shape as well as intensity (or absence) supports the helical structure of SWNTs.

## Methods

**Raman measurement**

We analyzed two kinds of commercial purified SWNTs, Tuball (Oscial, Russia) and ASA-100F (arc-discharge, Hanwha Chemical, South Korea), using Renishaw In-Via Raman Microscope. The measurements were carried out with laser excitation of 532 nm, spot size of 1-2 μm, laser power of 5mW and powder density of 1.6 mW/μm$^2$.

SWNTs samples were dehydrogenated in a high vacuum furnace (10$^{-5}$ Torr), kept at 600°C for 30 minutes.

**Raman simulation**

We carried out the simulation for regular (16, 16), (15, 15) and (10, 10) SWNTs. We also made the open structures both regular and distorted mimicking the helical structures (Fig. S3). The distorted SWNT can be thought of retaining multi curvatures with larger and smaller diameter than that of regular SWNT. The elliptic tubular structure and open curved graphene structures were also considered for simulations. We use density functional theory (DFT) as implemented in QUANTUM ESPRESSO simulation package.[41] Generalized gradient approximation (GGA) was used for exchange correlation energy of electrons and ultra-soft pseudopotentials to represent interaction between ionic cores and valence electrons.[42] Kohn Sham wave functions were represented with a plane wave basis with an energy cutoff of 40 Ry and charge density cutoff of 240 Ry. Integration over Brillouin zone (BZ) was sampled with a mesh of 1x1x2 grid.[43,44] Dynamical matrices at the Γ point (q=0) in BZ were computed using perturbative linear response approach used in DFT.[45]

**Supplementary information**

Supplementary Materials

**Figure captions**

Figure 1. Raman spectra of SWNT samples. (**a**) Raman spectra obtained from purified and bundled Ocsial- and arc-SWNTs samples revealing typical Raman signals for SWNTs. Inset shows the RBM (*l*-RBM) range of the spectra and a transmission electron microscope image for arc-CNT sample which appears as bundles. (**b**) Zoom-

in view of the spectra (a). Red numbers indicate Raman signals assigned to RTMs, which reveal the slowly varying profile with RBM and *G*.

Figure 2. Simulated Raman-active modes for different SWNT structures. (**a,a'**) Landau free energy landscape with one or two minima. (**b-d**) Raman modes simulated from regular (b) and opened tube structures (c) as well as curved graphene (d), revealing RBM, RTM and *G* in order. Diameters are ~2.2 nm for regular and opened tube structures. The blue numbers represent the number of nodes for each RTM. The red numbers in (b-d) represent frequencies of the modes. For curved graphene, each node reveals only one mode, due to lack of symmetry. (**e**) Diagrams explaining divergence and anti-symmetric curl of vector fields. (**f,g**) Schematic explaining Landau condition for RBM ($\delta>\delta_c$) (f) and RTM ($\delta<\delta_c$) (g).

Figure 3. Conceptual explanation of RTMs and variation of depth of focus with node. (**a,b**) 1$^{st}$ and 2$^{nd}$ node RTMs for the regular tube structure. The red solid dots indicate decentered focus of Eigen vectors of *l*-RM. $d_f$ is a depth of focus, the distance between the focus and the tube wall. Black dot in a,b indicates a focus of RBM where the depth of focus is *r*. (**a',b'**), Sine curves of the RTMs where the coherence length ($\xi$) is defined as a cycle of TM-RM-TM-RM-TM. Red and blue arrows indicate radial and transverse components deconvoluted from the slow varying Eigen vectors, respectively. Red and blue numbers in a,a'. (**b,b'**) indicate number of *l*-RM and TM, respectively. (**c**) Variation of the depth of the focus of *l*-RMs with a node. Inset shows a variation of the coherence length with a node.

Figure 4. Variation of RBM (*l*-RBM) and the 1$^{st}$ node RTM frequencies with the diameter of regular and opened SWNTs. Obtained diameters are ~1.4, ~2.0 and ~2.2 nm (Fig. S3).

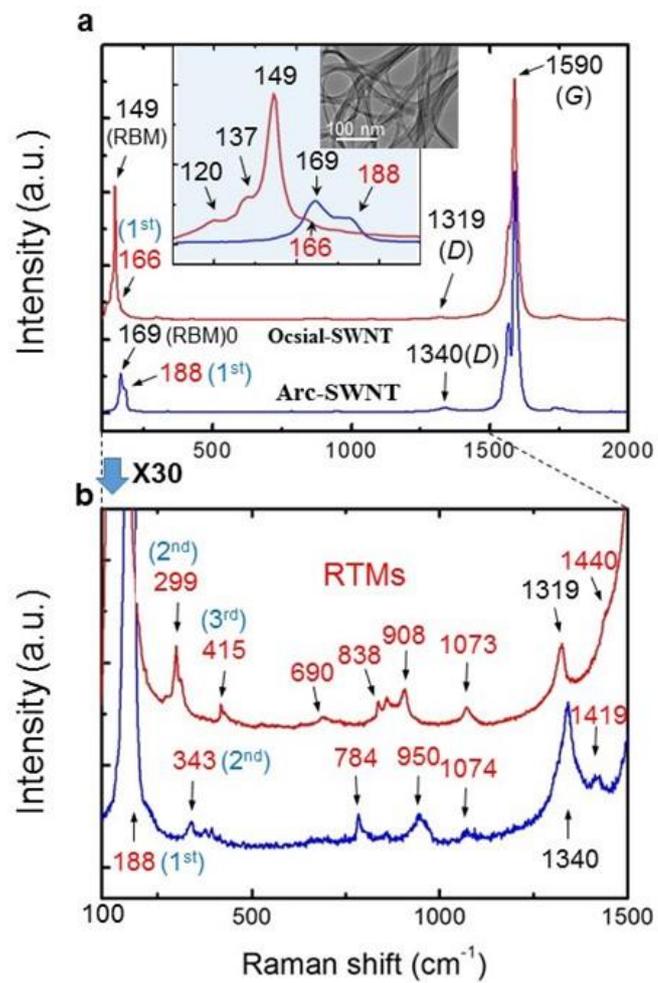

Fig. 1

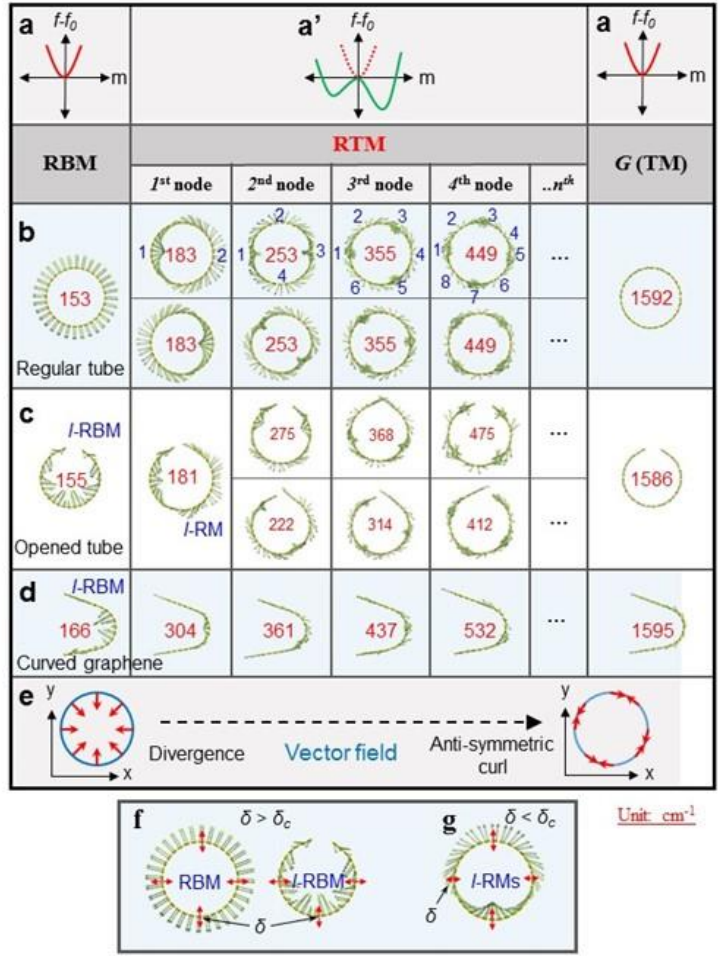

Fig. 2

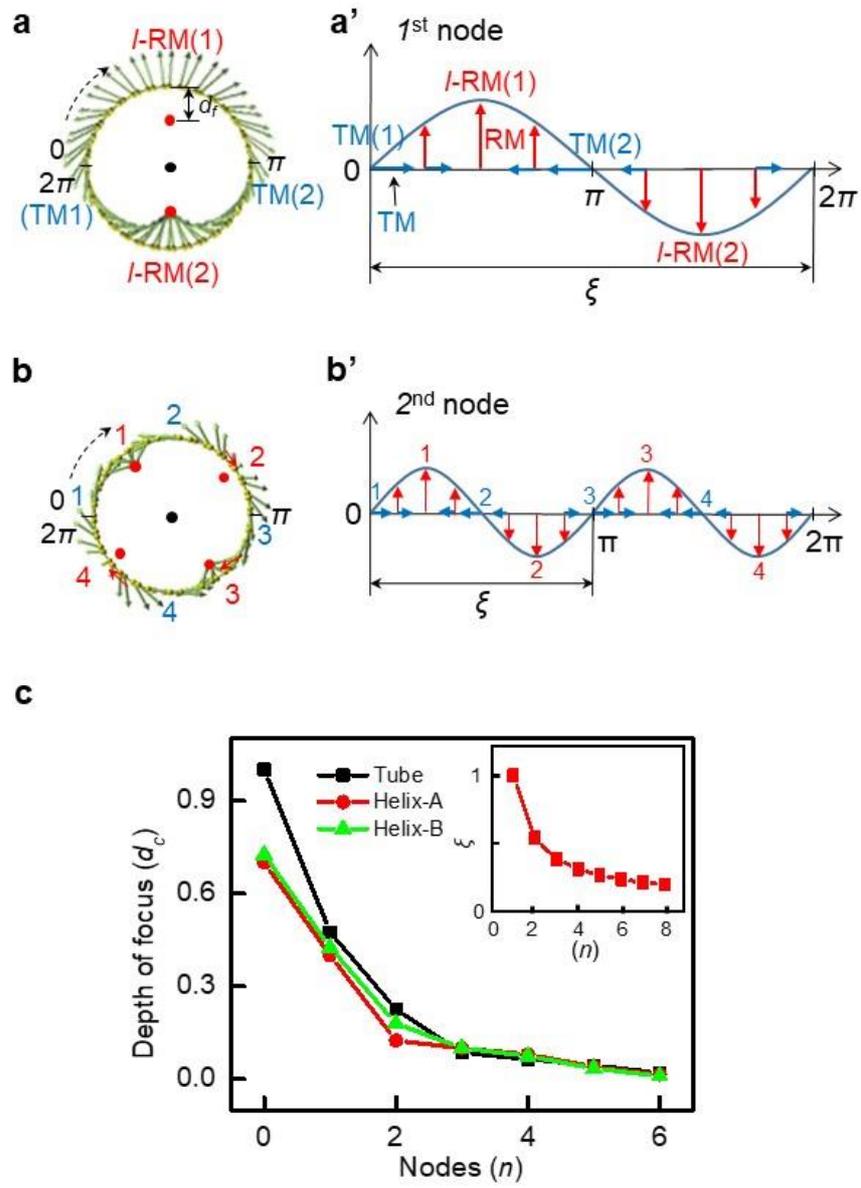

Fig. 3

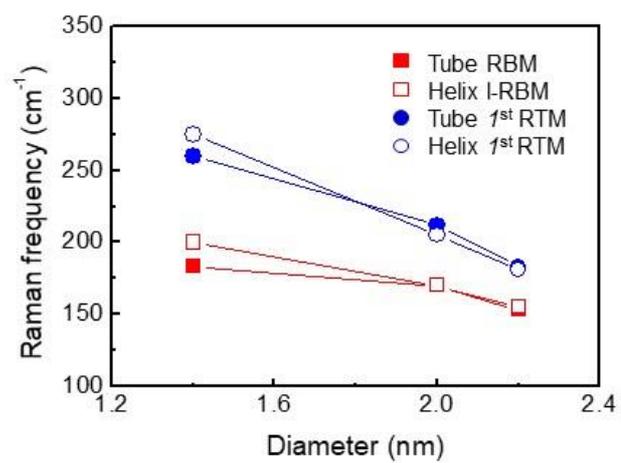

Fig. 4

*Supplementary Information for*

**Landau Raman Regulation Observed from Geometric Graphene Structures Including Single-Wall Carbon Nanotubes**

K. P. S. S. Hembram,[1] Jeongwon Park[2] and Jae-Kap Lee[1]*

[1]Center for Opto-Electronic Materials and Devices, Korea Institute of Science and Technology (KIST), Seoul 02792, Republic of Korea

[2]School of Electrical Engineering and Computer Science, University of Ottawa, Ottawa, Ontario, K1N 6N5, Canada

*Corresponding author: Email: jklee@kist.re.kr (Jae-Kap Lee)*

**This PDF file includes:**

- Supporting Information Figures S1 to S5
- Supporting Information text: Landau-Ginzburg theory for SWNT
- Supporting Information References

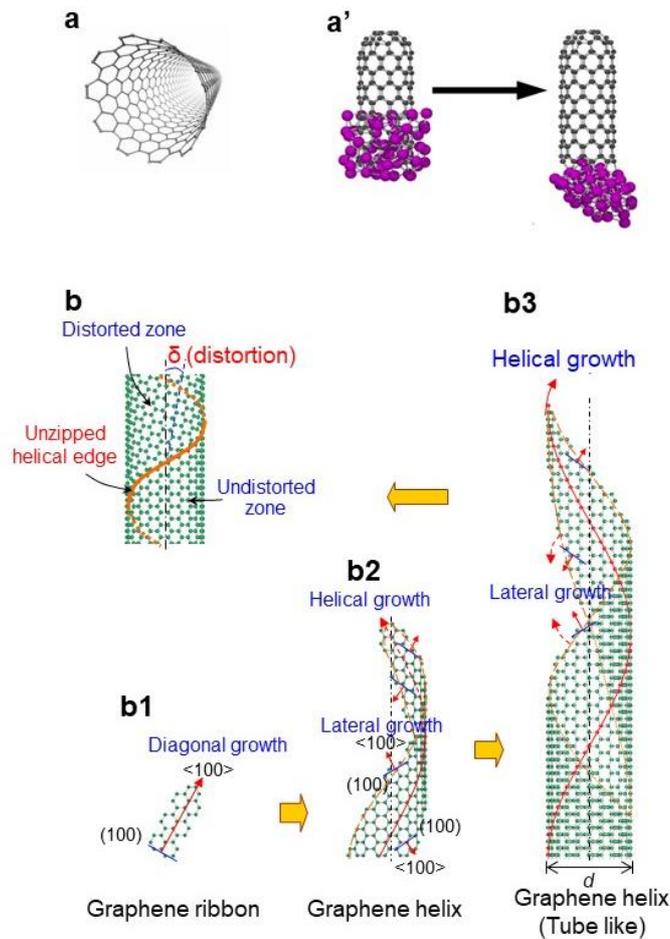

Figure S1. Two structure models for SWNTs. (a) Concentric tube model for SWNTs, resulted from tubular growth of SWNT nuclei (a'). This conventional model was derived from the structure model of multi-wall carbon nanotubes. (b) Helix model for SWNTs, resulted from helical growth of graphene nanoribbons (b1-b3). The model was suggested by Lee *et al.*[1] in 2014. In the helix model, a graphene nanoribbon is a nucleus (b1) which spirally and laterally grows resulting in tube-like graphene structures. The materials can be seen as perfect tubes, defective tubes and nodal tubes, according to the degree of the helical scroll, and generally appear to be locally distorted (see Fig. S2). In the helix model, chiral theory of SWNTs is not a necessary condition and the chirality is interpreted as distortion ($\delta$) (b).

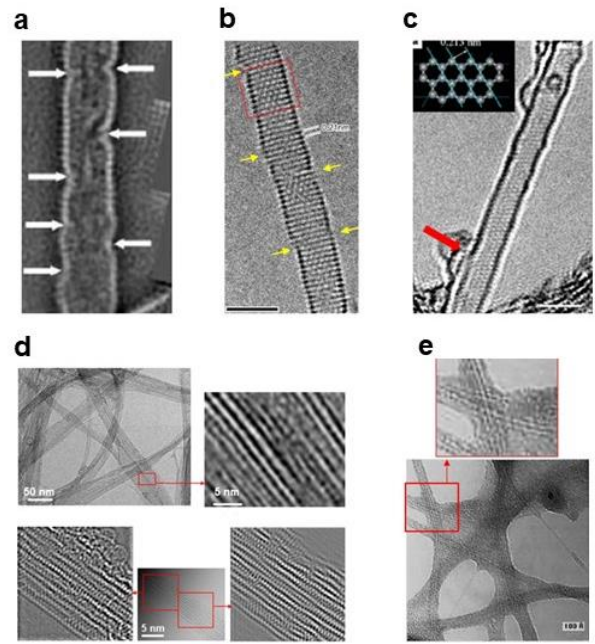

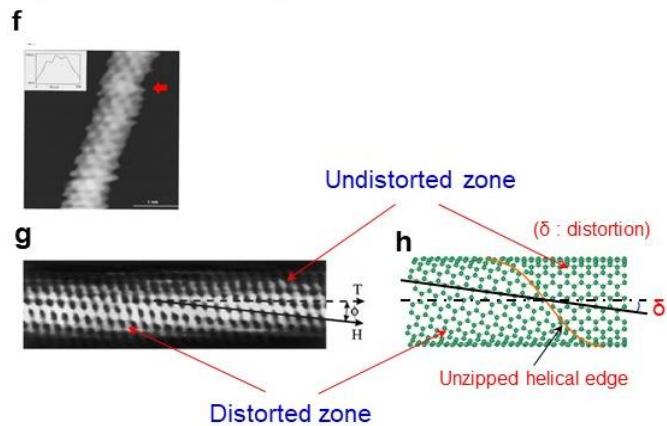

Figure S2. HRTEM and STM evidence for the helix model of SWNTs. (a) Meyer *et al.*'s HRTEM image,[2] revealing evidence of helical SWNT. They reported diameter of the samples used to be 1.46~1.63 nm. (b) Suenaga *et al.*'s HRTEM image.[3] Yellow arrows indicate the traces of helical SWNT. Scale bar=3 nm. (c) Hashimoto *et al.*'s HRTEM image.[4] The red arrow indicates disconnected lattice. The tubule is bumpy. These support helical hypothesis of SWNTs. Scale bar, 2 nm. (d) Lee *et al.*'s HRTEM images.[1] (e) Bethune *et al.*'s HRTEM image,[5] revealing nodal morphology which is evidence of the helical structure model of SWNTs. (f) Ge *et al.*'s STM image.[6] Red arrow indicate distorted zone of the tubule which is evidence of the helical structure model of SWNTs. (g) Wildöer, *et al.*'s STM image.[7] (h) New analysis for the STM image (g) in terms of the helix model. The chirality ($\varphi$) based on the concentric tube model (g) is interpreted as distortion $\delta$ (due to helical growth) in the helix model (h).

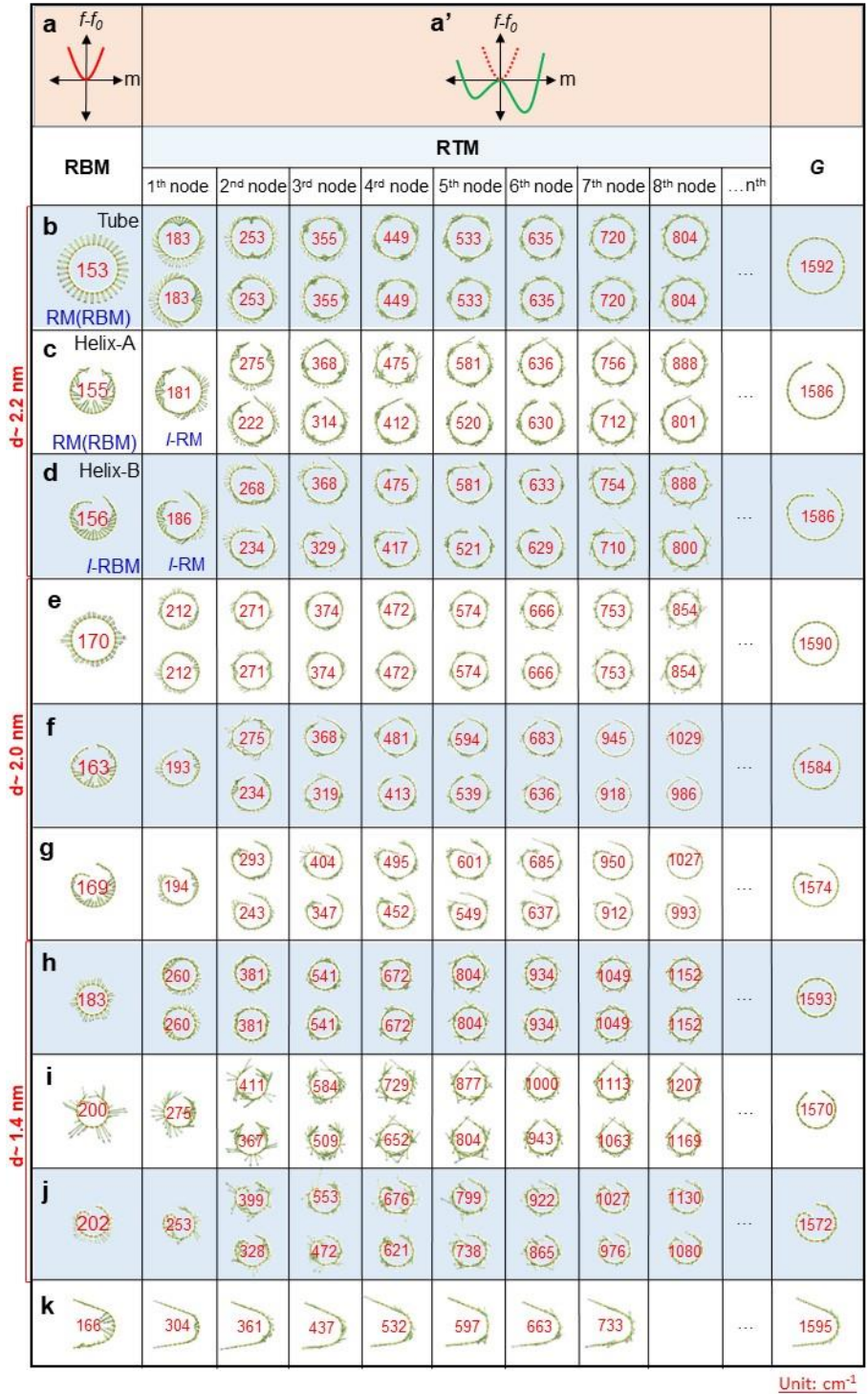

Figure S3. Simulated active-Raman spectra of different SWNT structures. (a,a') Landau free energy landscape. (b-d) Simulated active-Raman spectra for concentric tube (b) and opened (c,d) SWNTs with ~2.2 nm in diameter. (e-g) Simulated active-Raman spectra for concentric tube (e) and opened (f,g) SWNTs with ~2.0 nm in diameter. (h-j) Simulated active-Raman spectra for concentric tube (h) and opened (i,j) SWNTs with ~1.4 nm in diameter. (k) Simulated active-Raman spectra for curved graphene structure. The red numbers in (b-k) represent the frequency of the modes.

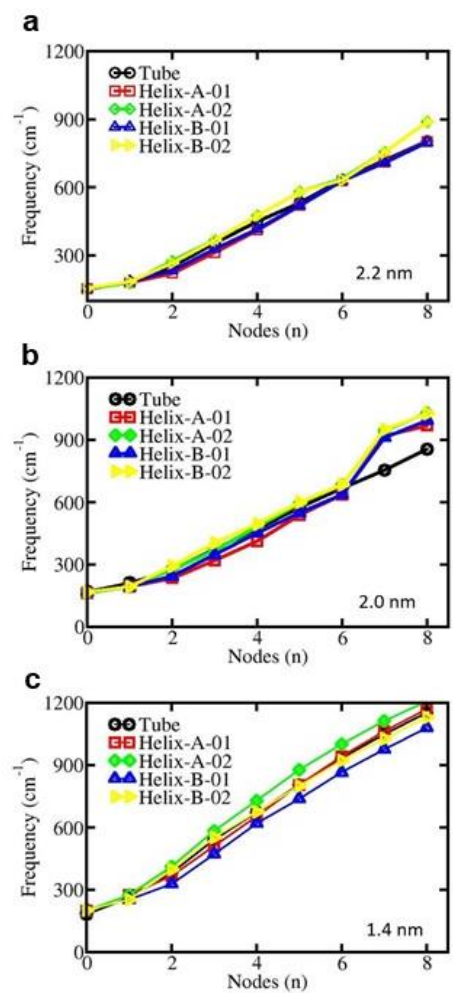

Figure S4. Variation of RBM and RTM frequency with nodes for concentric and opened (helical) SWNTs. Frequency of RBM (0[th] node) and RTMs is monotonically increase with nodes regardless of whether their structure is concentric or opened.

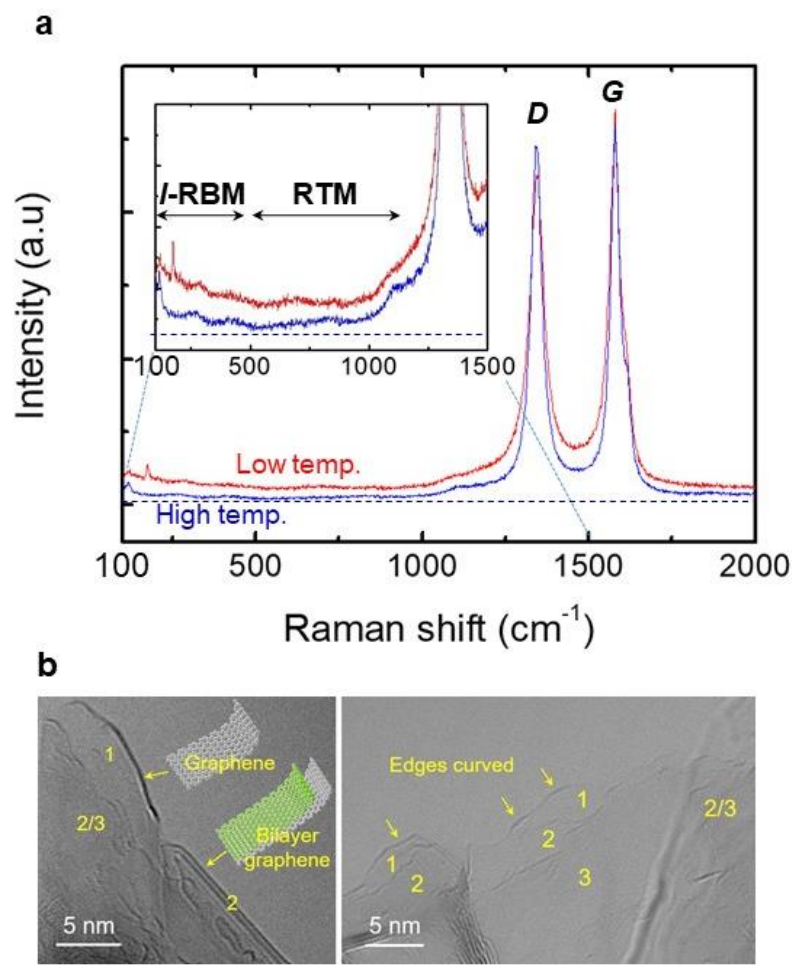

Figure S5. Raman spectra and TEM images of curved graphene. *l*-RBM and RTMs are evident in inset (a). Number of graphene layers can be identified from their curved edge structures (b) (Permission from ref.13).

## Landau-Ginzburg (*LG*) theory for SWNT:

We implement *LG* theory for SWNT to elucidate the helical geometry based on the spatial variation of radial mode (RM). In this context, we define RM which is an Eigen state to be the order parameter (OP) i.e. '*m*'. Expanding the free energy in terms of '*m*' we get $f = f_0 + \alpha m^2 + \frac{1}{2}\beta m^4 + higer\ terms$. Minimizing the free energy with respect to '*m*' we get $\frac{df}{dm} = 2m\alpha + 2\beta m^3 + ..$ Neglecting the higher order term and solving the equation, we obtain $m(\alpha + \beta m^2) = 0$. With $m = 0$, planar structure evolves. Hence no RM is observed for planar graphene. If $m \neq 0$, curvature exists with $(\alpha + \beta m^2) = 0$. Hence $m^2 = -(\alpha/\beta)$. So $f - f_0 = -(\alpha^2/2\beta)$.

## Spatial variation of radial mode:

Here we describe the development of transverse mode (TM) and explain the interface of RM and TM. Owing to the Schrodinger type equation we obtain,

$$-\frac{h^2}{2\mu}\frac{d^2 m}{dx^2} + \alpha m^2 + \frac{1}{2}\beta m^4 = 0$$

Parametrizing the order parameter '*m*'.

$$m_{rm} = m_{rbm} * u$$

$$\frac{dm_{rm}}{dx} = m_{rbm}\frac{du}{dx}$$

$$\frac{d^2 m_{rm}}{dx^2} = m^2{}_{rbm}\frac{d^2 u}{dx^2}$$

So $-\frac{h^2}{2\mu}m^2{}_{rbm}\frac{d^2 u}{dx^2} + \alpha\ m^2{}_{rbm}u^2 + \frac{1}{2}\beta\ m^4{}_{rbm}u^4$

Minimizing with respect to *u* we get

$$-\frac{h^2}{2\mu}\frac{d^2 u}{dx^2} + \alpha u + \beta u^3 = 0$$

Put $u = 1 + g$ where $g$ is the measure of order parameter

$$\frac{d^2 g}{dx^2} - \frac{2\mu g}{h^2}(\alpha + \beta g^2) = 0$$

$$\frac{d^2 g}{dx^2} - \frac{2g}{\xi^2} = 0$$

Solving the above equation, we get $g = exp\,(-x\sqrt{2})/\xi$, where $\xi$ is the characteristics coherence length.

With the variation of *m*, the higher nodes possess less amplitude and are hardly observed in the experiment (only RBM (or *l*-RBM) and $1^{st}$ RTM are observed).